\documentclass[prc,reprint,showpacs,amsmath]{revtex4-1}

\usepackage[T1]{fontenc}

\begin{document}

\title{Comment on ``No-core configuration-interaction model for the
  isospin- and angular-momentum-projected states'' by Satu\l a, %
  B\k aczyk, Dobaczewski, and Konieczka}

\author{I. Bentley}
\affiliation{Dept. of Physics, University of Notre Dame,
  Notre Dame, IN 46556}
\affiliation{Dept. of Chemistry and Physics, Saint Mary's College,
  Notre Dame, IN 46556}

\author{K. Neerg\aa rd}
\affiliation{Fjordtoften 17, 4700 N\ae stved, Denmark}

\author{S. Frauendorf}
\affiliation{Dept. of Physics, University of Notre Dame,
  Notre Dame, IN 46556}

\begin{abstract}
  The authors state that results recently published by Carlsson and
  Toivanen ``seem to contradict'' the conclusions of some of our work.
  We show that this statement is based on a misinterpretation of the
  results of Carlsson and Toivanen.
\end{abstract}

\maketitle

In Ref.~\cite{ref:Sa16}, Satu\l a, B\k aczyk, Dobaczewski, and
Konieczka write the following. (We have changed the reference numbers
in their text to correspond to the list of references at the end of
this comment.)

``It is worth recalling here that in the context of searching for
possible fingerprints of collective isoscalar $pn$-pairing phase in $N
\approx Z$ nuclei, the isoscalar pairing, or deuteron-like
correlations, were intensely discussed in the literature; see
Refs.~\cite{ref:Go72,ref:En96,ref:Satu97a,*ref:Satu01a,*ref:Satu01b}
and references cited therein. In particular, the isoscalar
$pn$-pairing was considered to be the source of an additional binding
energy that could offer a microscopic explanation of the so-called
Wigner energy~\cite{ref:Satu97b}---an extra binding energy along the
$N = Z$ line, which is absent in the self-consistent MF mass models.
In spite of numerous recent works following these early developments
attempting to explain the isoscalar $pn$-pairing correlations and the
Wigner energy (see
Refs.~\cite{ref:San12,ref:Be13,ref:Ca14,ref:Be14,ref:Sam15,ref:De15}
and references cited therein), the problem still lacks a satisfactory
solution.

There are at least two major reasons for that: (i) an incompleteness
of the HFB (HF) approaches used so far, which consider the $pn$ mixing
only in the particle-particle channel (see discussion in
Ref.~\cite{ref:Sato13}), and (ii) difficulties in evaluating the role
of beyond-mean-field correlations. Recently, within the RPA including
$pn$ correlations, the latter problem was addressed in
Ref.~\cite{ref:Ca14}. Their systematic study of the isoscalar and
isovector multiplets in magic and semi-magic nuclei rather clearly
indicated a missing relatively strong $T = 0$ pairing. This seems to
be in line with our NCCI model findings concerning description of %
$T = 0, I = 1$ states, but seems to contradict the conclusions of
Refs.~\cite{ref:Be13,ref:Be14}.''

Our reading of Ref.~\cite{ref:Ca14} by Carlsson and Toivanen does not
support that their findings ``seem to contradict'' our conclusions in
Refs.~\cite{ref:Be13,ref:Be14}. We model the masses of nuclei in the
range of mass numbers $A=24$--100 with no or little neutron excess. In
the doubly odd $N=Z$ nuclei we treat the lowest $T=0$ and $T=1$ states
separately, where $N$, $Z$, and $T$ denote the numbers of neutrons and
protons and the isospin. To expose the variation of the observed and
calculated masses with these variables we consider four different
combinations of the individual masses. In both works the model is
isobarically invariant except for a phenomenological Coulomb
contribution to the total mass.

In Ref.~\cite{ref:Be13} the independent nucleon plus isovector pairing
Hamiltonian is diagonalized exactly in a valence space formed by a
small number of Nilsson levels. The deformation of each nucleus is
taken from a previous calculation. A term proportional to $T(T+1)$ is
added to the calculated energies. It is concluded that this model
reproduces the variation with $A$ of the four mass combinations quite
well except that the symmetry energy coefficient is overestimated when
the coefficient of the $T(T+1)$ term is fit to the mass difference of
the lowest $T=1$ and $T=0$ states in the doubly odd $N=Z$ nuclei. A
Hamiltonian with a certain additional isoscalar pairing interaction
is also studied. A weak interaction of this type is found to have
little effect on the results of calculations while with a larger
coupling constant, the isoscalar pairing interaction destroys the
reproduction of the $N=Z$ doubly odd doubly even mass differences.

In Ref.~\cite{ref:Be14}, to allow for larger valence spaces, the exact
diagonalization of the isovector pairing Hamiltonian is replaced by
the Hartree-Bogolyubov plus random phase approximation (RPA). The
small valence spaces of Ref.~\cite{ref:Be13} are used to verify the
results based on this approach. Furthermore, a Strutinskij
renormalization is applied. The model may thus be described from
another point of view as a conventional Nilsson-Strutinskij
calculation amended by an RPA correction based on the same pairing
interaction as employed in the Bardeen-Cooper-Schrieffer term
traditionally included in such calculations. Enlarging the valence
space in this manner is found to eliminate the difficulty encountered
in Ref.~\cite{ref:Be13}, specifically that is reproducing the mass
difference of the lowest $T=1$ and $T=0$ states in the doubly odd
$N=Z$ nuclei and the symmetry energy coefficient simultaneously. The
112 masses of doubly even nuclei in our sample are reproduced with a
root mean square deviation of 0.95~MeV. Many features of the variation
with $A$ of the mass combinations are found explainable in terms of
the shell structure, that is, the pattern of Nilsson levels.
(Ref.~\cite{ref:Be13} also has a discussion of the impact of shell
structure on the so-called Wigner $X$. This is defined by a fit of the
Coulomb reduced masses of nuclei with equal $A$ and the lowest $T$ by
an expression proportional to $T(T+X)$ plus a constant.) The
approximate $T(T+1)$ form of the binding energies is understood as a
consequence of the spontaneous breaking of isospin symmetry by the
mean field. The isospin symmetry breaking is caused to equal parts by
the isovector pair field and the average potential, which is the
source of the phenomenological $T(T+1)$ term of our theory.

We disagree with the statement that the findings of other authors
``contradict'' the results of our calculations. Further, we find no
such statement in the article by Carlsson and Toivanen. In our
reading, these authors do not address the Wigner energy at all; no
mass is calculated. Their application of the RPA is to excitation
energies, specifically the energies of neutron-proton pair and neutron
hole-proton hole pair excitations of a doubly magic core. From the
outset they abstain from reproducing the observed absolute values of
these excitation energies, which would seem so constrain most directly
the interaction strength. They consider the relative energies of the
excited states with different angular momenta, which they fit by a
fairly schematic two-component interaction. A ratio of about 1.4 of
their isoscalar and isovector coupling constants is found to give the
best fit in neighbors of doubly magic nuclei with $N\ne Z$. When the
same interaction is applied to the neighbors of doubly magic nuclei
with $N=Z$ their RPA calculations give imaginary excitation energies.
The authors therefore dismiss these nuclei from their sample.

When an excitation energy calculated in the RPA is viewed in its
dependence on the interaction strength, before becoming imaginary it
must vanish. This means in the present context that the pair
separation energy equals minus the sum of the neutron and proton
chemical potentials of the core. The findings of Carlsson and Toivanen
therefore imply that, in their model, the $N=Z$ doubly odd doubly even
mass staggering vanishes in the neighborhoods of the doubly magic
nuclei. This is certainly very different from what is observed. Thus,
just contrary to contradicting our conclusions, the findings of
Carlsson and Toivanen concur with some of our conclusions in
Ref.~\cite{ref:Be13}: A strong isoscalar pairing interaction induces a
condensation of isoscalar pairs, which eliminates the $N=Z$ doubly odd
doubly even mass staggering.

The authors of Ref.~\cite{ref:Sa16} cite Ref.~\cite{ref:Satu97b} by
Satu\l a, Dean, Gary, Mizutory, and Nazarewicz. They fail to mention
our comment on this work in Ref.~\cite{ref:Be14}. There we present the
results of shell model calculation for $A=48$ similar to those of
Ref.~\cite{ref:Satu97b} albeit restricted to the valence space
including only the $1f_{7/2}$ shell. The valence space of
Ref.~\cite{ref:Satu97b} includes the shells $1f_{7/2}$, $2p_{3/2}$,
$1f_{5/2}$, and $2p_{1/2}$. Like the authors of
Ref.~\cite{ref:Satu97b} we find that the Wigner energy, defined as the
deviation of the Coulomb reduced $T=0$ mass from a quadratic fit to
the $T=2$ and $T=4$ masses, decreases drastically when the
interactions of isoscalar pairs are switched off. This turns out,
however, to result from a decrease of the total symmetry energy. The
Wigner $X$ simultaneously \emph{increases}. As $^{48}$Cr is the
central nucleus of the $1f_{7/2}$ shell, a similar analysis of the
masses calculated by Satu\l a, Dean, Gary, Mizutory, and Nazarewicz in
the larger valence space is not expected to lead to a qualitatively
different conclusion. A definite answer to this question awaits the
highly desirable publication of the individual calculated masses
whence the published combinations were extracted.

\bibliography{satula}

\end{document}